\documentclass[preprint,prd,aps,showpacs,showkeys,nofootinbib]{revtex4}
\usepackage{graphicx}
\usepackage{dcolumn}
\usepackage{bm}
\usepackage{color}
\usepackage[dvipsnames]{xcolor}
\usepackage{amssymb}
\usepackage{float}
\usepackage{subfigure}

\definecolor{black-blue}{RGB}{77,116,175}
\definecolor{black-yellow}{RGB}{231,162,33}
\definecolor{black-green}{RGB}{144,180,58}
\definecolor{black-red}{RGB}{246,95,50}

\begin{document}
\title{Neutrino transition magnetic moment in the $U(1)_X$SSM}
\author{Long Ruan$^{1,2,3}$, Shu-Min Zhao$^{1,2,3}$\footnote{zhaosm@mail.nankai.edu.cn}, Ming-Yue Liu$^{1,2,3}$, \nonumber\\
Xing-Yu Han$^{1,2,3}$, Xi Wang$^{1,2,3}$, Xing-Xing Dong$^{1,2,3}$\footnote{dongxx@hbu.edu.cn}}

\affiliation{$^1$ Department of Physics, Hebei University, Baoding 071002, China}
\affiliation{$^2$ Hebei Key Laboratory of High-precision Computation and Application of Quantum Field Theory, Baoding, 071002, China}
\affiliation{$^3$ Hebei Research Center of the Basic Discipline for Computational Physics, Baoding, 071002, China}

\date{\today}

\begin{abstract}
This paper investigates the neutrino transition magnetic moment in the $U(1)_X$SSM. $U(1)_X$SSM is the $U(1)$ extension of Minimal Supersymmetric Standard Model (MSSM) and its local gauge group is extended to $SU(3)_C\times SU(2)_L \times U(1)_Y\times U(1)_X$. To obtain this model, three singlet new Higgs superfields and right-handed neutrinos are added to the MSSM, which can explain the results of neutrino oscillation experiments. The neutrino transition magnetic moment is induced by electroweak radiative corrections. By applying effective Lagrangian method and on-shell scheme, we study the associated Feynman diagrams and the transition magnetic moment of neutrinos in the model. We fit experimental data for neutrino mass variances and mixing angles. Based on the range of data selection, the influences of different sensitive parameters on the results are analysed. The numerical analysis shows that many parameters have an effect on the neutrino transition magnetic moment, such as $g_X$, $M_2$, $\mu$, $\lambda_H$ and $g_{YX}$. For our numerical results, the order of magnitude of $\mu_{ij}^M/\mu_B$ is around $10^{-20}$ $\sim$ $10^{-19}$.
\end{abstract}

\keywords{transition magnetic moment, $U(1)_X$SSM, neutrino.}

\maketitle

\section{Introduction}

The Standard Model (SM) comes under the umbrella of quantum field theory, which describes the three main forces, the strong force, the weak force, and the electromagnetic force\cite{b0,b1,b2,b3}. Besides it predicts the existence of the Higgs. Although the SM has been a great success, its flaws are obvious. It doesn't explain the mass problem of neutrinos, the related issue of dark matter, and can't describe gravity\cite{n0,n1,n2}. Therefore, it must be extended. Scientists have made many extensions to the SM, among which the Minimal Supersymmetric Standard Model(MSSM) is a popular one. However, there are problems in MSSM such as the $\mu$-problem\cite{31} and massless neutrinos\cite{32}. To break through these problems, we use the $U(1)_X$SSM.
Under this model, we study neutrino transition magnetic moment. The study of it may indirectly lead to a new understanding of the neutrino properties and the mechanism of neutrino mass generation. Besides it may verify the correctness of $U(1)_X$SSM to some extent. It is also important in the long distance propagation of neutrinos in the magnetic fields of matter and vacuum\cite{13}. Previous research on neutrino transition magnetic moment includes analyses of Majorana neutrino effects on supernova neutrino oscillations\cite{111} and explanations of electron recoil anomalies\cite{112}. However, our work explores this phenomenon within a distinct model, aiming to contribute novel findings.

$U(1)_X$SSM is the extension of the MSSM with the $U(1)_X$ gauge group,
and the symmetry group is $SU(3)_C\times SU(2)_L \times U(1)_Y\times U(1)_X$\cite{7}. This extension adds three Higgs singlet superfields and right-handed neutrino superfields to the MSSM.
Consequently, there are five neutral CP-even Higgs component fields
($H_{u}^0,~H_{d}^0$, $\phi_{\eta}^0,~\phi_{\bar{\eta}}^0,~\phi_{S}^0$) in the model,
and mix together, forming a $5\times 5$ mass-squared matrix. Consequently, the mass of the lightest CP-even Higgs particle can be improved at the tree level. In the $U(1)_X$SSM, the small hierarchy problem in the MSSM is alleviated through the added right-handed neutrinos, sneutrinos, and extra Higgs singlets. The $\mu$-problem existing in the MSSM is relieved after the spontaneous symmetry breaking of the $S$ field in vacuum through $\lambda_H\hat{S}\hat{H}_u\hat{H}_d$. Through the term $Y_\nu\hat{\nu} \hat{l} \hat{H}_u$, the right-handed neutrinos and left-handed neutrinos mix together, which makes light neutrinos to obtain tiny masses through the seesaw mechanism. The existence of supersymmetry provides a very natural candidate for dark matter: neutralino. While, $SU(3)_C\times SU(2)_L \times U(1)_Y\times U(1)_X$ can provide several dark matter candidates: neutralino, sneutrino(CP-even, CP-odd) etc., and it protects the Higgs mass from radiative correction by the massive particles, which solves the gauge hierarchy. Under $U(1)_X$SSM, the transition magnetic moment of neutrino is induced by electroweak radiative corrections.

The previous studies\cite{13} have investigated the neutrino transition magnetic moment using the effective Lagrangian method and the mass-shell scheme, yielding reasonable numerical results. In this paper, a more comprehensive study of the neutrino transition magnetic moment at the $U(1)_X$SSM is presented. Using the effective Lagrangian method and the mass-shell scheme we obtain the expression for the neutrino transition magnetic moment. We derive the relevant Feynman diagrams and calculate the neutrino transition moment by combining the operators.
In numerical calculations, we perform neutrino mixing within experimentally constrained parameter ranges to determine viable parameter values. Additionally, we compare the effects of different reasonable parameters on transition magnetic moment and get the numerical results.

The paper is organized according to the following structure. In Sec.II, we mainly introduce the content of the
$U(1)_X$SSM including its superpotential, the general soft breaking terms, the mass matrices and couplings. In Sec.III, we give the analytical expressions of the transition magnetic moment about neutrino. In Sec.IV, we give the relevant parameters and numerical analysis. In Sec.V, we present a summary of this article. Some formulae are collected in the Appendix.

\section{The essential  content of  $U(1)_X$SSM}
$U(1)_X$SSM is the extension of MSSM and the local gauge group is $SU(3)_C\times SU(2)_L \times U(1)_Y\times U(1)_X$. $U(1)_X$SSM has new superfields, which include three Higgs singlets $\hat{\eta},~\hat{\bar{\eta}},~\hat{S}$, and right-handed neutrinos $\hat{\nu}_i$. The corresponding superpotential of the $U(1)_X$SSM is given by:
\begin{eqnarray}
&&W=l_W\hat{S}+\mu\hat{H}_u\hat{H}_d+M_S\hat{S}\hat{S}-Y_d\hat{d}\hat{q}\hat{H}_d-Y_e\hat{e}\hat{l}\hat{H}_d+\lambda_H\hat{S}\hat{H}_u\hat{H}_d
\nonumber\\&&\hspace{0.6cm}+\lambda_C\hat{S}\hat{\eta}\hat{\bar{\eta}}+\frac{\kappa}{3}\hat{S}\hat{S}\hat{S}+Y_u\hat{u}\hat{q}\hat{H}_u+Y_X\hat{\nu}\hat{\bar{\eta}}\hat{\nu}
+Y_\nu\hat{\nu}\hat{l}\hat{H}_u.
\end{eqnarray}
The two Higgs doublets and three Higgs singlets can be listed as follows:
\begin{eqnarray}
&&\hspace{1cm}H_{u}=\left(\begin{array}{c}H_{u}^+\\{1\over\sqrt{2}}\Big(v_{u}+H_{u}^0+iP_{u}^0\Big)\end{array}\right),
~~~~~~
H_{d}=\left(\begin{array}{c}{1\over\sqrt{2}}\Big(v_{d}+H_{d}^0+iP_{d}^0\Big)\\H_{d}^-\end{array}\right),
\nonumber\\&&\eta={1\over\sqrt{2}}\Big(v_{\eta}+\phi_{\eta}^0+iP_{\eta}^0\Big),~~~
\bar{\eta}={1\over\sqrt{2}}\Big(v_{\bar{\eta}}+\phi_{\bar{\eta}}^0+iP_{\bar{\eta}}^0\Big),~~
S={1\over\sqrt{2}}\Big(v_{S}+\phi_{S}^0+iP_{S}^0\Big).
\end{eqnarray}
And the vacuum expectation values of the states $H_u,H_d,\eta$, $\bar{\eta}$, $S$ are respectively $v_u, v_d$, $v_\eta$, $v_{\bar\eta}$ and $v_S$.
$H_u^+$ is the charged part of the doublet $H_u$. While, $H_u^0(P_u^0)$ is the neutral CP-even(CP-odd) part of $H_u$.
The similar condition is for the doublet $H_d$. $\phi_\eta^0(P_\eta^0)$ is the CP-even(CP-odd) part of singlet $\eta$.
$\phi_{\bar{\eta}}^0(P_{\bar{\eta}}^0)$ is the CP-even(CP-odd) part of singlet ${\bar{\eta}}$.
$\phi_S^0(P_S^0)$ is the CP-even(CP-odd) part of singlet $S$.

There are two angles defined as $\tan\beta=v_u/v_d$ and $\tan\beta_\eta=v_{\bar{\eta}}/v_{\eta}$. The soft SUSY breaking terms of $U(1)_X$SSM are shown as:
\begin{eqnarray}
&&\mathcal{L}_{soft}=\mathcal{L}_{soft}^{MSSM}-B_SS^2-L_SS-\frac{T_\kappa}{3}S^3-T_{\lambda_C}S\eta\bar{\eta}
+\epsilon_{ij}T_{\lambda_H}SH_d^iH_u^j\nonumber\\&&\hspace{1cm}
-T_X^{IJ}\bar{\eta}\tilde{\nu}_R^{*I}\tilde{\nu}_R^{*J}
+\epsilon_{ij}T^{IJ}_{\nu}H_u^i\tilde{\nu}_R^{I*}\tilde{l}_j^J
-m_{\eta}^2|\eta|^2-m_{\bar{\eta}}^2|\bar{\eta}|^2-m_S^2S^2\nonumber\\&&\hspace{1cm}
-(m_{\tilde{\nu}_R}^2)^{IJ}\tilde{\nu}_R^{I*}\tilde{\nu}_R^{J}
-\frac{1}{2}\Big(M_S\lambda^2_{\tilde{X}}+2M_{BB^\prime}\lambda_{\tilde{B}}\lambda_{\tilde{X}}\Big)+h.c~.
\end{eqnarray}
$\mathcal{L}_{soft}^{MSSM}$ represent the soft breaking terms of MSSM.
And $\lambda_{\tilde{B}}$ is the $U(1)_Y$ gaugino, which is the supersymmetric partener of the $U(1)_Y$ gauge boson $B^\mu$.
The boson of the added gauge group $U(1)_X$ is $X^\mu$, whose supersymmetric partener is $\lambda_{\tilde{X}}$.

\begin{table}[h]
\caption{ The superfields in $U(1)_X$SSM}
\begin{tabular}{|c|c|c|c|c|c|c|c|c|c|c|c|}
\hline
Superfields & $\hspace{0.1cm}\hat{q}_i\hspace{0.1cm}$ & $\hat{u}^c_i$ & $\hspace{0.2cm}\hat{d}^c_i\hspace{0.2cm}$ & $\hat{l}_i$ & $\hspace{0.2cm}\hat{e}^c_i\hspace{0.2cm}$ & $\hat{\nu}_i$ & $\hspace{0.1cm}\hat{H}_u\hspace{0.1cm}$ & $\hat{H}_d$ & $\hspace{0.2cm}\hat{\eta}\hspace{0.2cm}$ & $\hspace{0.2cm}\hat{\bar{\eta}}\hspace{0.2cm}$ & $\hspace{0.2cm}\hat{S}\hspace{0.2cm}$ \\
\hline
$SU(3)_C$ & 3 & $\bar{3}$ & $\bar{3}$ & 1 & 1 & 1 & 1 & 1 & 1 & 1 & 1  \\
\hline
$SU(2)_L$ & 2 & 1 & 1 & 2 & 1 & 1 & 2 & 2 & 1 & 1 & 1  \\
\hline
$U(1)_Y$ & 1/6 & -2/3 & 1/3 & -1/2 & 1 & 0 & 1/2 & -1/2 & 0 & 0 & 0  \\
\hline
$U(1)_X$ & 0 & -1/2 & 1/2 & 0 & 1/2 & -1/2 & 1/2 & -1/2 & -1 & 1 & 0  \\
\hline
\end{tabular}
\label{JJ1}
\end{table}

The particle content and charge assignments for $U(1)_X$SSM are mentioned in the table \ref {JJ1}. In our previous work, we have proven that $U(1)_X$SSM is anomaly free \cite{7}.

The covariant derivatives of $U(1)_X$SSM can be written as
\begin{eqnarray}
&&D_\mu=\partial_\mu-i\left(\begin{array}{cc}Y,&X\end{array}\right)
\left(\begin{array}{cc}g_{1},&g_{{YX}}\\0,&g_{{X}}\end{array}\right)
\left(\begin{array}{c}A_{\mu}^{Y} \\ A_{\mu}^{X}\end{array}\right)\;.
\end{eqnarray}
Compared with MSSM, $U(1)_X$SSM has a new effect called the gauge kinetic mixing, which is produced by Abelian groups $U(1)_Y$ and $U(1)_X$.
The basis conversion occurs when we use the rotation matrix $R$ ($R^TR = 1$), which is due to the fact that the two Abelian gauge groups are uninterrupted. The basis conversion can be described by \cite{UMSSM5,B-L1,B-L2,gaugemass}
\begin{eqnarray}
&&D_\mu=\partial_\mu-i\left(\begin{array}{cc}Y^Y,&Y^X\end{array}\right)
\left(\begin{array}{cc}g_{Y},&g{'}_{{YX}}\\g{'}_{{XY}},&g{'}_{{X}}\end{array}\right)R^TR
\left(\begin{array}{c}A_{\mu}^{\prime Y} \\ A_{\mu}^{\prime X}\end{array}\right)\;,\label{DMU}
\end{eqnarray}
with $A_{\mu}^{\prime Y}$ and $A_{\mu}^{\prime X}$ respectively representing the gauge fields of $U(1)_Y$ and $U(1)_X$.  Eq.~(\ref{DMU}) can be reduced to \cite{UMSSM5,B-L2,gaugemass}
\begin{eqnarray}
&&\left(\begin{array}{cc}g_{Y},&g{'}_{{YX}}\\g{'}_{{XY}},&g{'}_{{X}}\end{array}\right)
R^T=\left(\begin{array}{cc}g_{1},&g_{{YX}}\\0,&g_{{X}}\end{array}\right)~,~~~~
R\left(\begin{array}{c}A_{\mu}^{\prime Y} \\ A_{\mu}^{\prime X}\end{array}\right)
=\left(\begin{array}{c}A_{\mu}^{Y} \\ A_{\mu}^{X}\end{array}\right)\;.
\end{eqnarray}
Here $g_X$ expresses the gauge coupling constant of the $U(1) _X$ group and $g_{YX}$ expresses the mixing gauge coupling constant of the $U(1) _X$ and $U(1) _Y$ groups.

Some useful mass matrices and needed couplings in this model can be found in appendix A.

\section {formulation}
The magnetic dipole moment (MDM) and electric dipole moment (EDM) of the neutrino can actually be written as the operators
\begin{eqnarray}
&&\mathcal{L}_{MDM}=\frac{1}{2} \mu_{ij} \bar{\psi}_i \sigma^{\mu\nu} \psi_j F_{\mu\nu},\nonumber\\
&&\:\mathcal{L}_{EDM}=\frac{i}{2} \epsilon_{ij} \bar{\psi}_i \sigma^{\mu\nu} \gamma_5 \psi_j F_{\mu\nu},
\label{MEDM}
\end{eqnarray}
where $F_{\mu\nu}$ is the electromagnetic field strength, $\sigma^{\mu\nu}=\frac{i}{2}[\gamma^\mu,\gamma^\nu]$, $\psi_{i,j}$ denote the four-component Dirac fermions, $\mu_{ij}$ and $\epsilon_{ij}$ are respectively Dirac diagonal ($i=j$) or transition
($i\neq j$) MDM and EDM between states $\psi_{i}$ and $\psi_{j}$.

Since ${/\!\!\! p}=m_f \ll m_{_V}$ for on-shell fermions and ${/\!\!\! k}\rightarrow 0 \ll m_{_V}$ for photons, we can conveniently obtain the contribution of the loop diagram to the fermionic diagonal MDM and EDM using the effective Lagrangian method.  Then we can expand the amplitude of corresponding triangle diagrams based on the external momenta of fermion and photon. After matching the effective theory with the holonomic theory, we obtain all high-dimension operators along with their coefficients. We only need to keep those dimension 6 operators for later calculations.
\begin{eqnarray}
&&O_1^{L,R} = e \bar{\psi}_i {(i {/\!\!\!\! \mathcal{D}})}^3 P_{L,R} \psi_j , \nonumber\\
&&O_2^{L,R} = e \overline{(i \mathcal{D}_\mu {\psi}_i )} \gamma^\mu F\cdot \sigma P_{L,R} \psi_j, \nonumber\\
&&O_3^{L,R} = e \bar{\psi}_i  F\cdot \sigma \gamma^\mu P_{L,R} {(i \mathcal{D}_\mu {\psi}_j )}, \nonumber\\
&&O_4^{L,R} = e \bar{\psi}_i  (\partial^\mu F_{\mu\nu})  \gamma^\nu P_{L,R} \psi_j, \nonumber\\
&&O_5^{L,R} = e m_{{\psi}_i} \bar{\psi}_i  {(i {/\!\!\!\! \mathcal{D}})}^2 P_{L,R} \psi_j, \nonumber\\
&&O_6^{L,R} = e m_{{\psi}_i} \bar{\psi}_i F\cdot \sigma P_{L,R} \psi_j,
\label{operators}
\end{eqnarray}
where $P_L=\frac{1}{2}{(1 - {\gamma _5})}$, $P_R=\frac{1}{2}{(1 + {\gamma _5})}$, $\mathcal{D}_\mu=\partial^\mu+ieA_\mu$, and $m_{{\psi}_i}$ is the mass of fermion ${{\psi}_i}$.
The effective vertices with one external photon are written as
\begin{eqnarray}
&&O_1^{L,R} = ie \{[(p+k)^2+p^2]\gamma_\rho+({/\!\!\! p}+{/\!\!\! k})\gamma_\rho{/\!\!\! p}\} P_{L,R}, \nonumber\\
&&O_2^{L,R} = ie ({/\!\!\! p}+{/\!\!\! k})[{/\!\!\! k}, \gamma_\rho] P_{L,R}, \nonumber\\
&&O_3^{L,R} = ie [{/\!\!\! k}, \gamma_\rho] {/\!\!\! p} P_{L,R}, \nonumber\\
&&O_4^{L,R} = ie  (k^2\gamma_\rho-{/\!\!\! k}k_\rho) P_{L,R}, \nonumber\\
&&O_5^{L,R} = ie m_{{\psi}_i} \{({/\!\!\! p}+{/\!\!\! k})\gamma_\rho+\gamma_\rho {/\!\!\! p}\}  P_{L,R}, \nonumber\\
&&O_6^{L,R} = ie m_{{\psi}_i} [{/\!\!\!k}, \gamma_\rho]P_{L,R}.
\label{op}
\end{eqnarray}

By applying the equations of motion to the outer fermions, we obtain the relations in the effective Lagrangian\cite{Feng}:
\begin{eqnarray}
&&\quad\; C_2^{R}O_2^R + C_2^{L}O_2^L +  C_2^{L\ast}O_3^R + C_2^{R\ast}O_3^L + C_6^{R}O_6^R + C_6^{R\ast}O_6^L \nonumber\\
&&\Rightarrow (C_2^{R} + \frac{m_{{\psi}_j}}{m_{{\psi}_i}}C_2^{L\ast} + C_6^{R})O_6^R + (C_2^{R\ast} + \frac{m_{{\psi}_j}}{m_{{\psi}_i}}C_2^{L} + C_6^{R\ast})O_6^L \nonumber\\
&&=e m_{{\psi}_i} \Re(C_2^{R} + \frac{m_{{\psi}_j}}{m_{{\psi}_i}}C_2^{L\ast} + C_6^{R})\bar{\psi}_i \sigma^{\mu\nu} \psi_j  F_{\mu\nu} \nonumber\\
&&\quad +\: ie m_{{\psi}_i}\Im(C_2^{R} + \frac{m_{{\psi}_j}}{m_{{\psi}_i}}C_2^{L\ast} + C_6^{R})\bar{\psi}_i \sigma^{\mu\nu} \gamma_5 \psi_j   F_{\mu\nu},
\label{CtoRI}
\end{eqnarray}
Comparing with Eq.~(\ref{MEDM}) and Eq.~(\ref{CtoRI}), we can obtain
\begin{eqnarray}
&&\mu_{ij} = 4m_e m_{{\psi}_i} \Re(C_2^{R} + \frac{m_{{\psi}_j}}{m_{{\psi}_i}}C_2^{L\ast} + C_6^{R}) \mu_{\rm{B}},\nonumber\\
&&\:\epsilon_{ij} = 4m_e m_{{\psi}_i} \Im(C_2^{R} + \frac{m_{{\psi}_j}}{m_{{\psi}_i}}C_2^{L\ast} + C_6^{R}) \mu_{\rm{B}},
\end{eqnarray}
where $\Re(\cdots)$ and $\Im(\cdots)$ are the real and imaginary parts of the complex numbers respectively, $\mu_{\rm{B}}=e/(2 m_e)$, and $m_e$ is the electron mass. The Wilson coefficients$(C_2^{R},C_2^{L},C_6^{R},C_6^{L})$ related to our study in this paper have been included in the Appendix B.

\begin{figure}[ht]
\setlength{\unitlength}{4.0mm}
\centering
\includegraphics[scale=0.5]{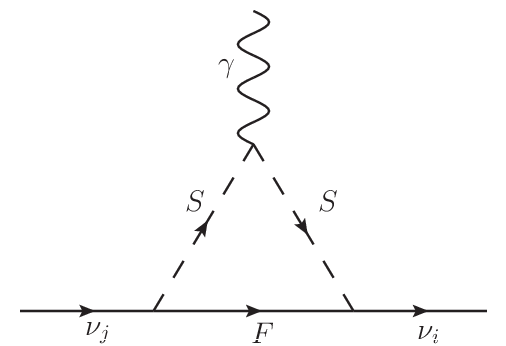}
\setlength{\unitlength}{4.0mm}
\centering
\includegraphics[scale=0.5]{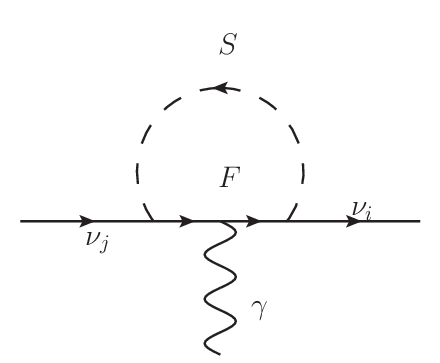}
\caption{Feynman diagrams for the $\nu_j\rightarrow{\nu_i\gamma}$ processes in the $U(1)_X$SSM.}\label{N1}
\end{figure}

Then we investigate the $\nu_j\rightarrow \nu_i \gamma$ processes about the transition magnetic moment of neutrino under the $U(1)_X$SSM. The amplitude of $\nu_j\rightarrow \nu_i\gamma$ can be obtained from the following Feynman diagrams shown in Fig.\ref{N1}.
After calculating the left one in Fig.(\ref{N1}) and connecting with Eq.(\ref{op}), we can get

\begin{eqnarray}
&&\mathcal M=-i\int\frac{d^Dk}{(2\pi)^D}\frac{1}{(k^2-m_F^2)(k^2-m_S^2)^2}
\Big\{-\frac{1}{4}(\frac{k^2}{k^2-m_S^2}-\frac{k^4}{(k^2-m_S^2)^2})\nonumber\\&&\times(O_2^{L}+O_3^{L})A_RB_L-
\frac{1}{2}(1-\frac{k^2}{k^2-m_S^2})O_6^{L}A_LB_L\Big\},
\label{mu}
\end{eqnarray}

where $k$ is the photon momentum, $m_F$ corresponds to the chargino mass, $m_S$ corresponds to the scalar lepton  mass. $A_L,~A_R$,~$B_R$ and~$B_L$ are
\begin{eqnarray}
&&A_L=-g_2U_{j1}^*\sum_{a=1}^{3}U_{ia}^{V,*}Z_{ka}^{E}+U_{j2}^*\sum_{a=1}^{3}U_{ia}^{V,*}Y_{e,a}Z_{k,(3+a)}^{E},\nonumber\\
&&A_R=\sum_{a=1}^{3}Y_{\nu,a}^{*}U_{i,(3+a)}^VZ_{ka}^{E}V_{j2},\nonumber\\
&&B_R=A_L^*,\nonumber\\
&&B_L=A_R^*.
\end{eqnarray}
Through the general description of the electromagnetic form factors of Dirac and Majorana neutrinos, we can get the MDM and EDM for Majorana neutrinos
\begin{eqnarray}
&&\mu_{ij}^M = \mu_{ij}^D -\mu_{ji}^D,\qquad\epsilon_{ij}^M = \epsilon_{ij}^D -\epsilon_{ji}^D.
\end{eqnarray}
Finally, we simplify Eq.(\ref{mu}) and use numerical calculation software(Mathematica) to get the numerical results.

\section{Numerical analysis}

In this section of the numerical results, we consider some constraints from experiments, including:

1. The lightest CP-even Higgs $h^0$ mass is around 125.1 GeV. The Higgs $h^0$ decays ($h^0\rightarrow \gamma+\gamma,~ Z+Z,~ W+W,~ b+\bar{b},~\tau+\bar{\tau}$) can well meet the latest experimental constraints\cite{18,su1,su2}.

The mass of the lightest CP-even Higgs boson should consider the stop quark contributions at loop level\cite{1251,1252}
\begin{eqnarray}
&&m_h^0=\sqrt{(m_{h_1}^0)^2+\Delta m_h^2},\label{higgs mass}
\end{eqnarray}
with $m_{h_1}^0$ representing the lightest tree-level Higgs boson mass. The concrete form of $\Delta m_h^2$ is
\begin{eqnarray}
&&\Delta m_h^2=\frac{3m_t^4}{4\pi^2 v^2}\Big[\Big(\tilde{t}+\frac{1}{2}\tilde{X}_t\Big)+\frac{1}{16\pi^2}\Big(\frac{3m_t^2}{2v^2}-32\pi\alpha_3\Big)\Big(\tilde{t}^2
+\tilde{X}_t \tilde{t}\Big)\Big],\nonumber\\
&&\tilde{t}=\log\frac{M_{\tilde{T}}^2}{m_t^2},\qquad\;\tilde{X}_t=\frac{2\tilde{A}_t^2}{M_{\tilde{T}}^2}\Big(1-\frac{\tilde{A}_t^2}{12M_{\tilde{T}}^2}\Big).\label{higgs corrections}\label{mnu}
\end{eqnarray}
$\alpha_3$ is the strong coupling constant. $M_{\tilde{T}}=\sqrt{m_{\tilde t_1}m_{\tilde t_2}}$
and $m_{\tilde t_{1,2}}$ are the stop masses. $\tilde{A}_t=A_t-\mu \cot\beta$ and $A_t$ is the trilinear Higgs stop coupling.
We use the parameter values to fix $m_h^0\sim 125.1 {\rm GeV}$.

The mass matrix of chargino includes the parameters $v_u,~v_d,~\lambda_H,~ v_S$, and the mass squared matrix of scalar lepton includes $v_u,~v_d,~\lambda_H,~ v_S,~ v_\eta,~v_{\bar{\eta}},~g_X,~g_{YX}$. The CP-even Higgs mass squared matrix at tree level also has these parameters.  As we fit CP-even Higgs $h^0$ mass, these parameters will be restricted, which affects the values of these parameters.
In the whole, these parameters affect the mass matrix of chargino and the mass squared matrix of scalar lepton. Therefore, they have effects on the neutrino transition magnetic moment.

2. The constraints from neutrino experiment data including mixing angles and mass variances are considered\cite{1,2,3}.

3. The $Z^\prime$ boson mass is larger than 5.1 TeV.
The gauge boson masses are\cite{7}
\begin{eqnarray}
&&\qquad\;\quad\;M_\gamma^2=0,\nonumber\\
&&\qquad\;\quad\;M_{Z,{Z^{'}}}^2=\frac{1}{8}\Big((g_{1}^2+g_2^2+g_{YX}^2)v^2+4g_{X}^2\xi^2 \nonumber\\
&&\qquad\;\qquad\;\qquad\;\mp\sqrt{(g_{1}^2+g_{2}^2+g_{YX}^2)^2v^4+8(g_{YX}^2-g_{1}^2-
g_{2}^2)g_{X}^2v^2\xi^2+16g_{X}^4\xi^4}\Big).
\end{eqnarray}
For $M_{Z^\prime}$, it can be much simplified with the supposition $\xi^2\gg v^2$.
It is shown as
\begin{eqnarray}
&&M_{Z^{'}}^2=\frac{1}{8}\Big((g_{1}^2+g_2^2+g_{YX}^2)v^2+4g_{X}^2\xi^2 \nonumber\\
&&+\sqrt{(g_{1}^2+g_{2}^2+g_{YX}^2)^2v^4+8(g_{YX}^2-g_{1}^2-
g_{2}^2)g_{X}^2v^2\xi^2+16g_{X}^4\xi^4}\Big)
\nonumber\\&&\approx\frac{1}{8}\Big(4g_{X}^2\xi^2 +\sqrt{16g_{X}^4\xi^4}\Big)
\nonumber\\&&=g^2_X\xi^2.
\end{eqnarray}
$M_{Z^\prime}/g_X > 6$ TeV is the results of the particle\cite{21}.

4. The neutralino mass is limited to more than 116 GeV, and the chargino mass is limited to more than 1000 GeV. The slepton mass is limited to more than 600 GeV \cite{18}.

With the above experimental requirements, we generally take the values of new  mass parameters($M_{BB'}$,$M_{BL}$,$M_S$) around the energy scale of new physics($10^3$ GeV). $M_{\tilde{L}}$,$M_{\tilde{E}}$ are all of mass square dimension, and can
be up to the order of $10^6$ GeV$^2$. Non-diagonal elements of scalar lepton mass matrix affect $T_e$, which can reach $10^{-1}$ GeV. $\tan\beta$ and $v_S$ affect the mass matrix of chargino. $v_{\bar{\eta}}$ and $v_{\eta}$ affect slepton mass.
The loop diagram is produced by chargino and scalar lepton.
We adopt the following parameters that can affect the neutrino transition magnetic moment in the numerical calculation:
\begin{eqnarray}
&&\tan\beta=23,~~~v_S=4.3~{\rm TeV},~~~ \tan\beta_\eta=\frac{v_{\bar{\eta}}}{v_{\eta}}=0.8,
\nonumber\\&& v_{\bar{\eta}}=17\sin(\beta_\eta)~{\rm TeV},~~~ v_{\eta}=17\cos(\beta_\eta)~{\rm TeV},
\nonumber\\&&T_{e11} = T_{e22} = T_{e33} = 0.5~{\rm GeV},
\nonumber\\&&M_{\tilde{L}11} = M_{\tilde{L}22} = M_{\tilde{L}33} = 3~\rm{TeV}^2,
\nonumber\\&&M_{\tilde{E}11} =M_{\tilde{E}22} =M_{\tilde{E}33} = 8~\rm{TeV}^2.
\end{eqnarray}

And the parameters we selected are of good universality.
In the following numerical analysis, the parameters to be studied include:
\begin{eqnarray}
g_X,~~~\lambda_H,~~~ M_2,~~~ \mu,~~~ g_{YX}.
\end{eqnarray}
Without special statement, the non-diagonal elements of the parameters are supposed as zero.

\subsection{ Neutrino Mixing}
In the neutrino mass matrix, elements such as $Y_{\nu}$ are relevant to neutrino mixing. Transition magnetic moment is closely related to the mass matrix including $Y_{\nu}$.
In this subsection, using the top-down approach we can derive the formulae for the neutrino mass and mixing angle from the effective neutrino mass matrix. Here, we adopt the normal ordering spectrum to calculate the neutrino observables($\sin^2(\theta_{ij}$) etc.). The detailed procedure is outlined in Appendix C.

The constraints from neutrino experiment data are\cite{18}
\begin{eqnarray}
&&\sin^2(\theta_{12})=0.307^{+0.013}_{-0.012},\nonumber\\
&&\sin^2(\theta_{23})=0.546\pm0.021,\nonumber\\
&&\sin^2(\theta_{13})=0.022\pm0.0007,\nonumber\\
&&\Delta{m^2_\odot}=(7.53\pm0.18)\times10^{-5}~{\rm eV}^2,\nonumber\\
&&|\Delta{m_{A}^2}|=(2.453\pm0.033)\times10^{-3}~{\rm eV}^2.
{\label{E6}}
\end{eqnarray}

To fit the data of neutrino physics, we take the parameters as
\begin{eqnarray}
&&Y_{X11}=Y_{X22}=Y_{X33}=0.1,~~~Y_{\nu_{22}} = 1.4000\times10^{-6},\nonumber\\
&&Y_{\nu_{33}} = 1.352420\times10^{-6},~~~~~~~~Y_{\nu_{12}} = 7.604202\times10^{-8}.\label{Ynu}
\end{eqnarray}
By fixing some matrix elements in Eq.(\ref{Ynu})
 and taking others as variables, we can discuss data easier.

\begin{figure}[H]
\setlength{\unitlength}{5mm}
\centering
\includegraphics[width=3.0in]{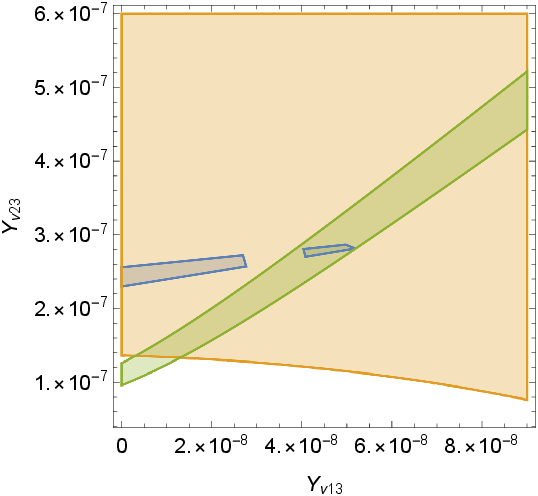}
\vspace{0.2cm}
\caption{$\sin^2(\theta_{12})$, $\sin^2(\theta_{23})$ and 10 $\sin^2(\theta_{13})$ are plotted in the plane of $Y_{\nu13}$ versus $Y_{\nu23}$. The yellow, blue and green area represent $0.483<\sin^2(\theta_{23})<0.609$, $0.271<\sin^2(\theta_{12})<0.346$ and $0.199<10\sin^2(\theta_{13})<0.241$ respectively.}{\label {N2}}
\end{figure}

In Fig.\ref{N2}, $\sin^2(\theta_{12})$, $\sin^2(\theta_{23})$ and 10 $\sin^2(\theta_{13})$ are plotted in the plane of $Y_{\nu23}$ versus $Y_{\nu13}$. If the area satisfies the 10 $\sin^2(\theta_{13})$ in the $30\sigma$, it of course satisfies $\sin^2(\theta_{23})$ in the $3\sigma$. With $Y_{\nu_{11}} = 1.092847\times10^{-6}$, the constraints of three mixing angles are satisfied(They all in the range of $3\sigma$). The yellow, blue and green area represent $0.483<\sin^2(\theta_{23})<0.609$, $0.271<\sin^2(\theta_{12})<0.346$ and $0.199<10\sin^2(\theta_{13})<0.241$ respectively. The yellow region resembles a rectangle, the blue region a fragmented ribbon, and the green region a continuous ribbon. The overlapping area represents values of $Y_{\nu23}$ and $Y_{\nu13}$ that satisfy all three mixing angle constraints.

\begin{figure}[H]
\setlength{\unitlength}{5mm}
\centering
\includegraphics[width=3.0in]{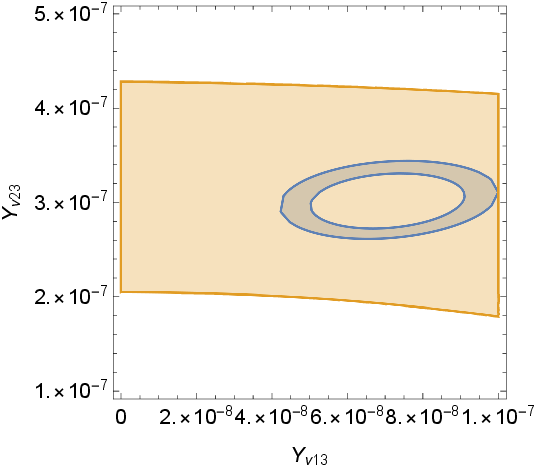}
\vspace{0.2cm}
\caption{$\Delta{m^2_\odot}$ and $|\Delta{m_{A}^2}|$ are plotted in the plane of $Y_{\nu13}$ versus $Y_{\nu23}$. The yellow area represents $1.2\times10^{-21}~{\rm eV}^2<|\Delta{m_{A}^2}|<4.3\times10^{-21}~{\rm eV}^2$ and the blue area represents $7.3\times10^{-23}~{\rm eV}^2<\Delta{m^2_\odot}<9.9\times10^{-23}~{\rm eV}^2$.}{\label {N3}}
\end{figure}

Similarly, in Fig.\ref{N3} the three constraints from the mixing angles are satisfied(They are all in the range of $3\sigma$). $\Delta{m^2_\odot}$ and $|\Delta{m_{A}^2}|$ are plotted in the plane of $Y_{\nu23}$ versus $Y_{\nu13}$. The yellow area represents $2.353\times10^{-21}~{\rm eV}^2<|\Delta{m_{A}^2}|<2.553\times10^{-21}~{\rm eV}^2$, which looks like a rectangle. The blue area represents $6.99\times10^{-23}~{\rm eV}^2<\Delta{m^2_\odot}<8.07\times10^{-23}~{\rm eV}^2$, which looks like a band. Overall the overlapping part is needed.

\begin{figure}[H]
\setlength{\unitlength}{5mm}
\centering
\includegraphics[width=3.0in]{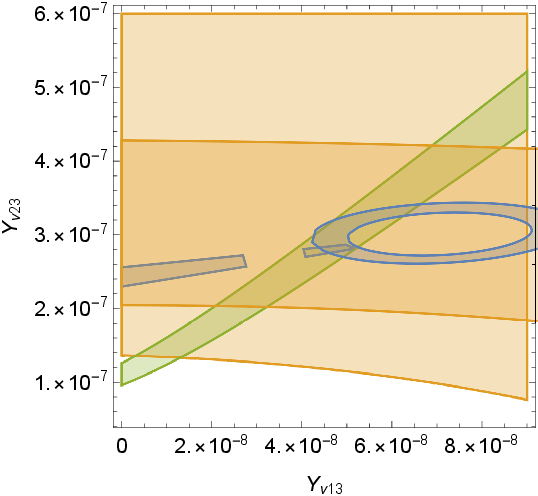}
\vspace{0.2cm}
\caption{Combining Fig.\ref{N2} and Fig.\ref{N3}, the overlapping area satisfy all constraints.}{\label {N4}}
\end{figure}

In Fig.\ref{N4}, we combine Fig.\ref{N2} and Fig.\ref{N3} to find reasonable parameter space apparently. The overlapping area in Fig.\ref{N2} satisfies $0.483<\sin^2(\theta_{23})<0.609$, $0.271<\sin^2(\theta_{12})<0.346$ and $0.199<10\sin^2(\theta_{13})<0.241$. In Fig.\ref{N3}, the overlapping area satisfies $1.2\times10^{-21}~{\rm eV}^2<|\Delta{m_{A}^2}|<4.3\times10^{-21}~{\rm eV}^2$ and $7.3\times10^{-23}~{\rm eV}^2<\Delta{m^2_\odot}<9.9\times10^{-23}~{\rm eV}^2$. In this figure, all shadow overlapping areas meet five constraints.

\begin{figure}[ht]
\setlength{\unitlength}{5mm}
\centering
\includegraphics[width=3.0in]{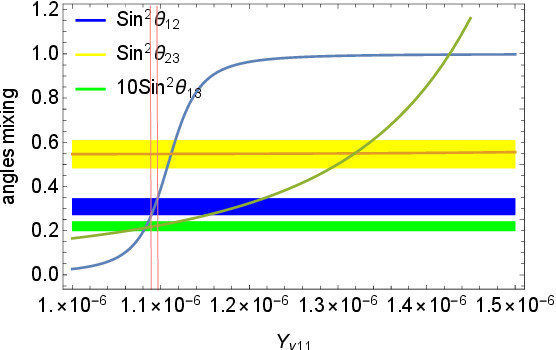}
\vspace{0.2cm}
\caption{$\sin^2(\theta_{12})$, $\sin^2(\theta_{23})$ and 10$\times$$\sin^2(\theta_{13})$ are plotted as $Y_{\nu11}$ changes. The blue line represents $\sin^2(\theta_{12})$. The yellow line represents $\sin^2(\theta_{23})$. The green line represents 10$\times$$\sin^2(\theta_{13})$. The blue, yellow and green regions correspond to the values of the $\sin^2(\theta_{12})$, $\sin^2(\theta_{23})$ and 10$\times$$\sin^2(\theta_{13})$ mixing angles in the $3\sigma$ range. The pink lines represent the overlapping satisfaction interval.}{\label {N5}}
\end{figure}

Now we discuss how the matrix element $Y_{\nu}$ such as $Y_{\nu_{11}}$ affects the $\sin^2(\theta_{12})$, $\sin^2(\theta_{23})$ and 10$\times$$\sin^2(\theta_{13})$. In Fig.\ref{N5}, the constraints from two mass variances are satisfied. Then $\sin^2(\theta_{12})$, $\sin^2(\theta_{23})$ and 10$\times$$\sin^2(\theta_{13})$ are plotted as $Y_{\nu11}$ changes. According to the analysis of Fig.\ref{N4}, we can take $Y_{\nu_{13}} = 4.516926\times10^{-8}$ and $Y_{\nu_{23}} = 2.803229\times10^{-7}$. With the above data, the blue, yellow and green regions correspond to the values of $\sin^2(\theta_{12})$, $\sin^2(\theta_{23})$ and 10$\times$$\sin^2(\theta_{13})$ mixing angles in the $3\sigma$ range, respectively. The blue line represents $\sin^2(\theta_{12})$. It grows consistently from $Y_{\nu_{11}}=1.0\times10^{-6}$ to $Y_{\nu_{11}}=1.3\times10^{-6}$, with rapid growth from $Y_{\nu_{11}}=1.07\times10^{-6}$ to $Y_{\nu_{11}}=1.136\times10^{-6}$,
but it remains almost constant in the $Y_{\nu_{11}}$ region $[1.3\times10^{-6},~ 1.5\times10^{-6}]$. The yellow line represents $\sin^2(\theta_{23})$.
We can find that it keeps stable going from $Y_{\nu_{11}}=1.0\times10^{-6}$ to $Y_{\nu_{11}}=1.5\times10^{-6}$, which is always in the range of $3\sigma$. The green line represents 10$\times$$\sin^2(\theta_{13})$. With the increase of $Y_{\nu_{11}}$, the 10$\times$$\sin^2(\theta_{13})$ grows faster and faster.
From that all, we can see that in order to satisfy the mixing angle from the experiments $Y_{\nu11}$ must be taken between two pink lines. So $Y_{\nu11}$ should be one of the values in the range from $1.08902\times10^{-6}$ to $1.9701\times10^{-6}$.

Using the Gaussian likelihood function, we construct a function combining three mixing angles and two mass variances:
\begin{eqnarray}
p(\mathbf{y}) = \prod_{i=1}^5 \frac{1}{\sqrt{2\pi\sigma^2_i}} \exp\left(-\frac{(y_i - \mu_i)^2}{2\sigma^2_i}\right),
\label{1234}
\end{eqnarray}
where $y_i$ and $\mu_i$ from 1 to 5 represent three mixing angles and two mass variances respectively.
We take
\begin{eqnarray}
&&\mu_1=\sin^2(\theta_{12})=0.307,\nonumber\\
&&\mu_2=\sin^2(\theta_{23})=0.546,\nonumber\\
&&\mu_3=\sin^2(\theta_{13})=0.022,\nonumber\\
&&\mu_4=\Delta{m^2_\odot}=7.53\times10^{-5}~{\rm eV}^2,\nonumber\\
&&\mu_5=|\Delta{m_{A}^2}|=2.453\times10^{-3}~{\rm eV}^2.
\label{8888}
\end{eqnarray}
And $\sigma_i$ correspond to their standard deviation.
\begin{figure}[H]
\setlength{\unitlength}{5mm}
\centering
\includegraphics[width=3.0in]{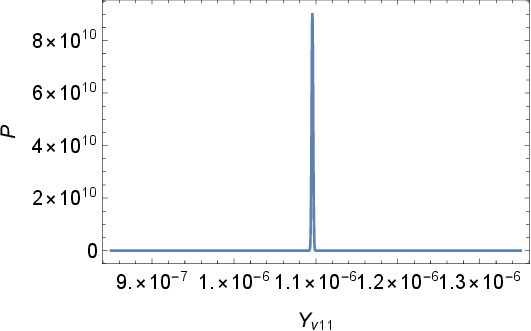}
\vspace{0.2cm}
\caption{The relationship between $Y_{\nu11}$ and $P$.}{\label {M5}}
\end{figure}
The extreme value of the ordinate in Fig.\ref{M5} corresponds to the value of our parameter. We can get the value of $Y_{\nu_{11}}$= $1.092847\times10^{-6}$.

In summary, combining the five experimental constraints on neutrinos, we can determine the range of values for our selected parameters. Through our analysis, we can determine that the following parameters are reasonable:
\begin{eqnarray}
&&Y_{\nu_{11}}=1.092847\times10^{-6},~Y_{\nu_{22}} = 1.4000\times10^{-6},~Y_{\nu_{33}} = 1.352420\times10^{-6},\nonumber\\
&&Y_{\nu_{12}} = 7.604202\times10^{-8},~Y_{\nu_{13}} = 4.516926\times10^{-8},~Y_{\nu_{23}} = 2.803229\times10^{-7}.\label{MMM}
\end{eqnarray}

\subsection{The processes of $\nu_j\rightarrow{\nu_i\gamma}$}
In this part, the objective of this study is to investigate the influence of certain sensitive parameters
on the numerical results of neutrino transition magnetic moment $\mu_{ij}^M$ under experimental constraints. In the following discussion, we use the Eq.~(\ref{MMM}) to continue the numerical calculation. Besides, $\mu_{ij}^M$ is used to represent the transition magnetic moment of the Majorana neutrinos.
We choose a number of parameters and investigate them to the extent allowed, such as $g_X$, $\lambda_H$, $M_2$, $\mu$.

\begin{figure}[H]
\setlength{\unitlength}{5mm}
\centering
\includegraphics[width=3in]{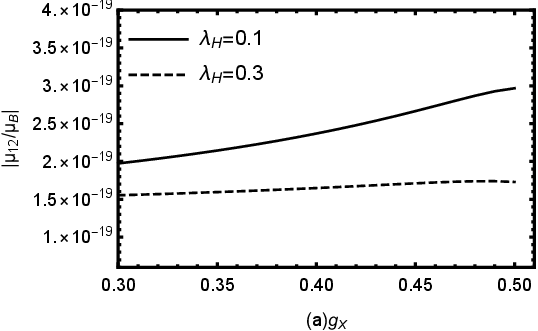}
\vspace{0.2cm}
\setlength{\unitlength}{5mm}
\centering
\includegraphics[width=3.15in]{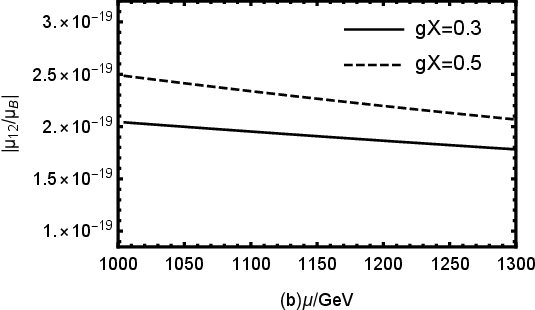}
\setlength{\unitlength}{5mm}
\centering
\includegraphics[width=3.1in]{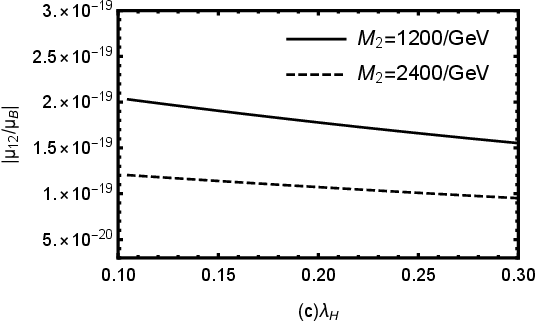}
\setlength{\unitlength}{5mm}
\centering
\includegraphics[width=3.15in]{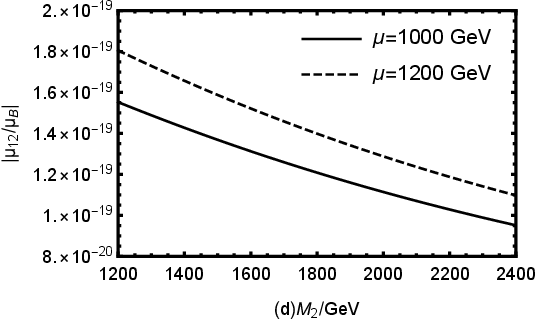}
\setlength{\unitlength}{5mm}
\centering
\caption{The relationship between different parameters and $\mu_{12}^M$/$\mu_B$.}{\label {2}}
\end{figure}
$g_X$ is the gauge coupling constant of the new gauge group $U(1)_X$. Besides, the mass matrixes of slepton and coupling vertices $\nu_i\chi_j^-\tilde{e}_k^{*}$ all have the important parameter $g_X$, which can improve the new physics effect.
We plot $g_X$ and $\mu_{12}^M/\mu_B$ in the Fig.\ref{2}~(a), in which the dashed line corresponds to $\lambda_H$ = 0.3 and the solid line corresponds to $\lambda_H$ = 0.1. Here, we take $\mu$ = 1000~GeV and $M_2$ = 1200~GeV.
We find that the both lines increase in the most region of $g_X$ during the range of 0.3-0.51. And the solid line is larger than the dashed line. Generally speaking, larger $g_X$ should lead to larger UMSSM contributions.

In Fig.\ref{2}~(b), with $g_X$ = 0.5(dashed line) and $g_X$ = 0.3(solid line), we take $\lambda_H$ = 0.1 and $M_2$ = 1200~GeV. As solid and dashed lines go from bottom to top, $\mu_{12}^M/\mu_B$ increases as $g_X$ increases. They are decreasing functions of $\mu$. $\mu$ appears in the term $\frac{1}{\sqrt{2}}$$\lambda_{H}$$v_S$+$\mu$ in the mass matrix for the chargino, which may has influence on the result. As shown in Fig.\ref{2}~(b), with the increase of $\mu$, the chargino mass becomes heavier, which suppresses the numerical results.

$\lambda_H$ comes from the term $\lambda_H\hat{S}\hat{H}_u\hat{H}_d$ in the superpotential. The mass matrices of several particles(chargino, neutralino) all have the important parameter $\lambda_H$, which possibly produces complex effects on the numerical results. In Fig.\ref{2}~(c), we take $\mu$ = 1000~GeV and $g_X$ = 0.3. The solid and dashed lines respectively represent $M_2 = 1200$~GeV and $M_2 = 2400$~GeV. Both the dashed and solid lines are decreasing functions as $\lambda_H$ turns large.

In Fig.\ref{2}~(d), we take $\lambda_H$ = 0.1 and $g_X$=0.3. The solid and dashed lines represent $\mu = 1000$~GeV and $\mu = 1200$~GeV respectively. Similarly, $M_2$ as the mass matrix element of chargino,
has the similar effect to $\mu$ on $\mu_{12}^M/\mu_B$. We also can see that $\mu_{12}^M/\mu_B$ decreases as $M_2$ increases.

The above discussion is about $\mu_{12}^M/\mu_B$. For $\mu_{13}^M$/$\mu_B$ and $\mu_{23}^M$/$\mu_B$,
the influence of certain sensitive parameters are very similar as the condition of $\mu_{12}^M/\mu_B$. Therefore, we only list some of the parameters and plot their effects.

\begin{figure}[H]
\includegraphics[width=3.15in]{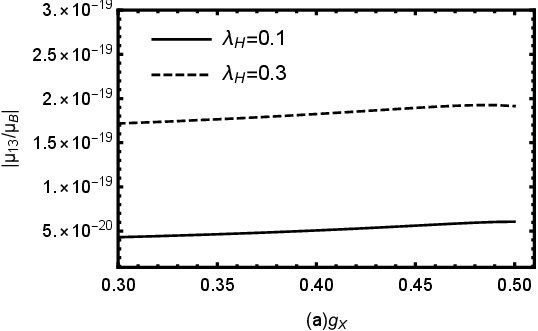}
\setlength{\unitlength}{5mm}
\centering
\includegraphics[width=3.15in]{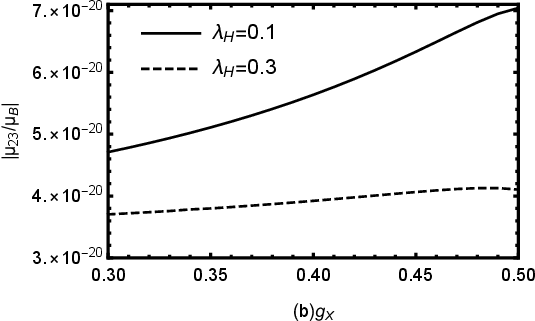}
\setlength{\unitlength}{5mm}
\centering
\includegraphics[width=3.15in]{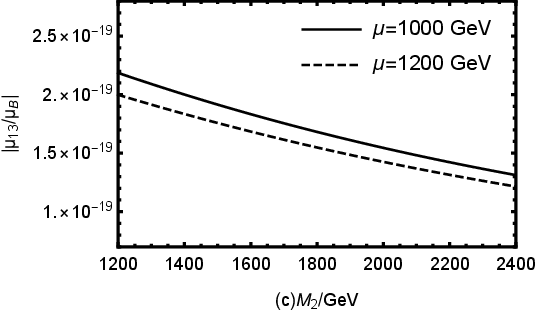}
\setlength{\unitlength}{5mm}
\centering
\includegraphics[width=3.15in]{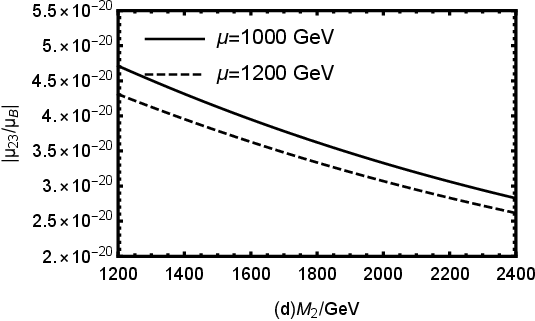}
\caption{The relationship between different parameters with $\mu_{13}^M$/$\mu_B$ and $\mu_{23}^M$/$\mu_B$.}{\label {222}}
\end{figure}

Fig.\ref{222}~(a) and Fig.\ref{222}~(b) describe the relationship between $g_X$ and $\mu_{13}^M/\mu_B$ and $\mu_{23}^M$/$\mu_B$. Like the description of $\mu_{13}^M/\mu_B$, the dashed line corresponds to $\lambda_H$ = 0.3 and the solid line corresponds to $\lambda_H$ = 0.1. With $\mu$ = 1000~GeV and $M_2$ = 1200~GeV, we can find that the effects of $g_X$ on the different components of $\mu_{ij}^M$/$\mu_B$ have similar trends. For Fig.\ref{222}~(a), within the value range, the maximum values for solid and dashed lines are respectively $3.27\times10^{-19}$ and $1.92\times10^{-19}$. For Fig.\ref{222}~(b), the maximum values for solid and dashed lines are $7.04\times10^{-20}$ and $4.12\times10^{-20}$.

Fig.\ref{222}~(c) and Fig.\ref{222}~(d) describe the influences of $M_2$ on $\mu_{13}^M/\mu_B$ and $\mu_{23}^M/\mu_B$. We can also find the similar trends, and they are decreasing functions of $M_2$.
In Fig.\ref{222}~(c), the maximum values for the solid and dashed lines are $1.69\times10^{-19}$ and $1.55\times10^{-19}$ respectively. The maximum values for the dashed and solid lines in the Fig.\ref{222}~(d) are $3.10\times10^{-19}$ and $4.03\times10^{-20}$.

From the above graphs we can conclude that $\mu_{ij}^M$/$\mu_B$ increase with $g_X$ and decrease with $\mu$, $\lambda_H$ and $M_2$. Their influences on $\mu_{ij}^M$/$\mu_B$ tend to be similar.
Overall, $g_X$, $\lambda_H$, $\mu$ and $M_2$ are indeed sensitive parameters that have obvious impacts on $\mu_{ij}^M$/$\mu_B$.

To explore the $\mu_{ij}^M$ parameter spaces well, we plot scatter diagrams for several parameters shown in Fig.\ref{3}.
The scanned parameters are listed in TABLE \ref{c1}. Then, we use $\textcolor{brown}{\blacksquare}~(\mu_{12}^M/\mu_B<1.6\times 10^{-19}),~\textcolor{Purple}{\blacklozenge}~(1.6\times 10^{-19}\leq \mu_{12}^M/\mu_B<1.9\times 10^{-19}),~ \textcolor{Red}{\blacktriangle}~(1.9\times 10^{-19}\leq \mu_{12}^M/\mu_B<2.2\times 10^{-19}),~\textcolor{Blue}{\bullet}~2.2\times 10^{-19}\leq \mu_{12}^M/\mu_B<3.2\times 10^{-19})$ to represent the results of the transition magnetic moment.

\begin{table*}[h]
\caption{Scanning parameters for Fig.\ref{3}}
\begin{tabular*}{\textwidth}{@{\extracolsep{\fill}}|l|l|l|l|@{}}
\hline
Parameters&$g_X~~~~~~~~~~~~~~~~$&$g_{YX}~~~~~~~~~~~~~~~$&$M_2/{\rm GeV}$~~~~~~~~~~~~\\
\hline
Min&0.3&0.01&~~~1000\\
\hline
Max&0.4&0.3&~~~2000\\
\hline
\end{tabular*}
\label{c1}
\end{table*}

\begin{figure}[H]
\setlength{\unitlength}{5mm}
\centering
\includegraphics[width=3.15in]{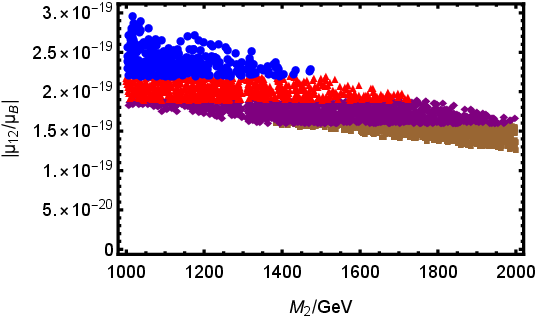}
\setlength{\unitlength}{6mm}
\centering
\includegraphics[width=3.15in]{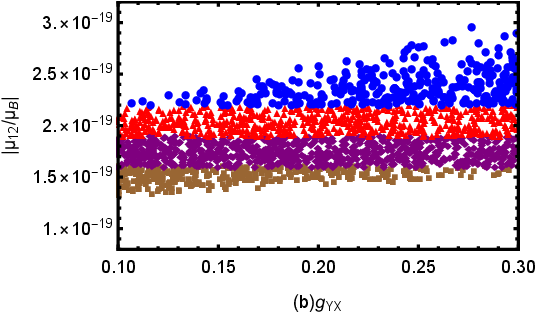}
\setlength{\unitlength}{5mm}
\vspace{0.1cm}\\
\centering
\includegraphics[width=3.15in]{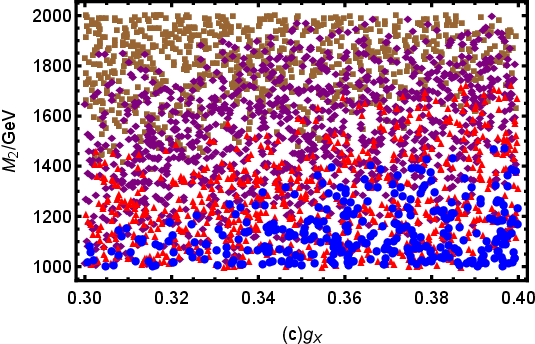}
\setlength{\unitlength}{5mm}
\centering
\includegraphics[width=3.15in]{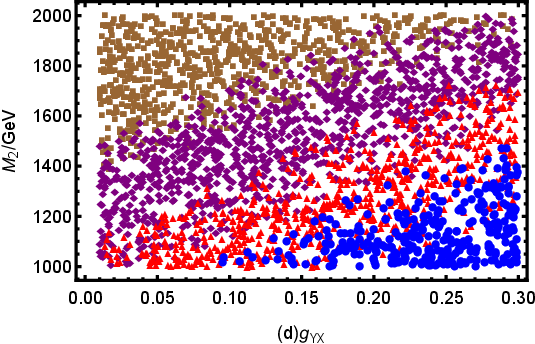}
\setlength{\unitlength}{5mm}
\vspace{0.1cm}\\
\centering
\includegraphics[width=3.15in]{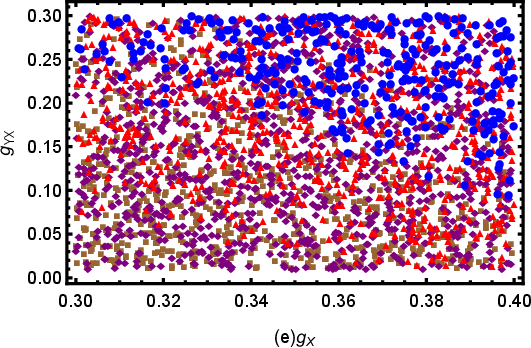}
\caption{The relationship between different parameters and $\mu_{12}^M/\mu_B$.}{\label {3}}
\end{figure}

In Fig.\ref{3}(a), scatter plots of $M_2$ versus $\mu_{12}^M/\mu_B$ are shown. The overall figure resembles a parallelogram. We can see that $\textcolor{Blue}{\bullet}$ is at the top, $\textcolor{Red}{\blacktriangle}$ and $\textcolor{Purple}{\blacklozenge}$ are in the middle and $\textcolor{Red}{\blacktriangle}$ is on top of $\textcolor{Purple}{\blacklozenge}$. Finally $\textcolor{brown}{\blacksquare}$ is at the bottom. It can be seen that $\mu_{12}^M/\mu_B$ decreases as $M_2$ increases. We can also see from the graph that its maximum value is $3.0\times 10^{-19}$. This result is consistent with the result of line $\mu_{12}^M/\mu_B$.
In Fig.\ref{3}(b), scatter plots of $g_{YX}$ versus $\mu_{12}^M/\mu_B$ are shown. It has the same graphic color layout as Fig.\ref{3}(a), just with a different trend. It can be gotten that the value of $\mu_{12}^M/\mu_B$ increases as $g_{YX}$ increases.

\begin{figure}[H]
\setlength{\unitlength}{5mm}
\centering
\includegraphics[width=3.15in]{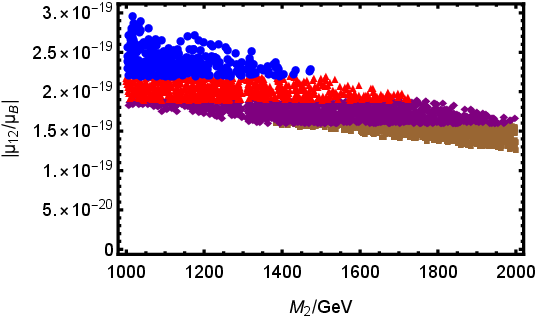}
\setlength{\unitlength}{6mm}
\centering
\includegraphics[width=3.15in]{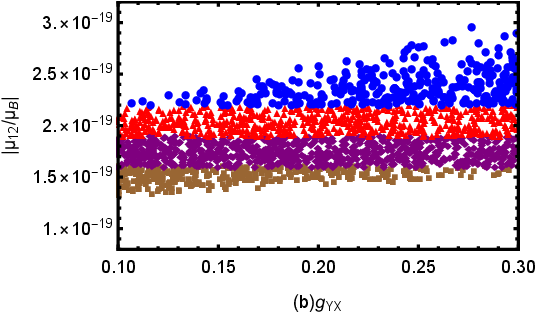}
\setlength{\unitlength}{5mm}
\vspace{0.1cm}\\
\centering
\includegraphics[width=3.15in]{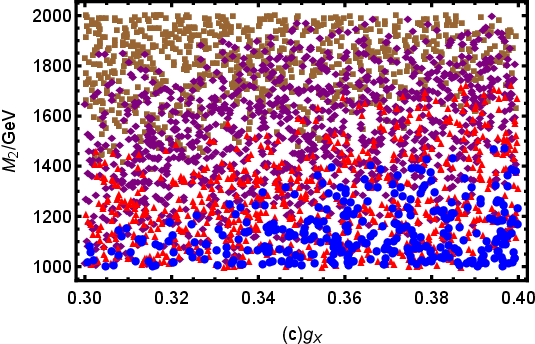}
\setlength{\unitlength}{5mm}
\centering
\includegraphics[width=3.15in]{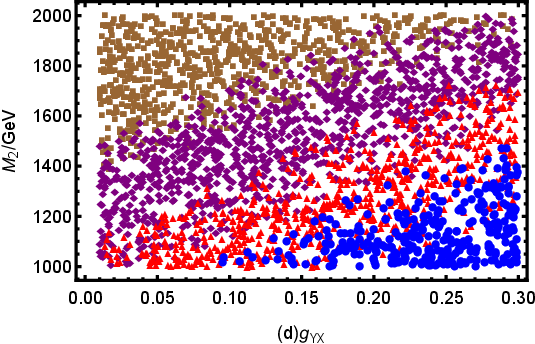}
\setlength{\unitlength}{5mm}
\vspace{0.1cm}\\
\centering
\includegraphics[width=3.15in]{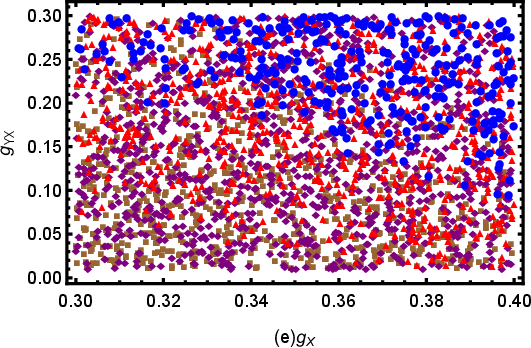}
\caption{The relationship between different parameters and $\mu_{12}^M/\mu_B$.}{\label {3}}
\end{figure}

Fig.\ref{3}(c) shows the effects of $g_{X}$ and $M_2$ on $\mu_{12}^M/\mu_B$. Viewed from the whole, different values of $\mu_{12}^M/\mu_B$ in the parameter space have obvious stratification. The upper left corner is $\textcolor{brown}{\blacksquare}$, followed by $\textcolor{Purple}{\blacklozenge}$, immediately followed by $\textcolor{Red}{\blacktriangle}$, and to the bottom right by $\textcolor{Blue}{\bullet}$. From the trend in the graph, we get that $\mu_{12}^M/\mu_B$ reaches its maximum value when $g_{X}$ = 0.4 and $M_2$ = 1000 within the parameter space of Fig.\ref{3}(c).
In Fig.\ref{3}(d), its graphical distribution is similar to that of Fig.\ref{3}(c). It proves that the effects of $g_{YX}$ and $g_X$ on $\mu_{12}^M/\mu_B$ are similar.

From Fig.\ref{3}(e), we can derive the effects of $g_{X}$ and $g_{YX}$ on $\mu_{12}^M/\mu_B$. The color distribution is obvious from the overall view, the upper right corner is a mix of blue, red and purple colors with some brown. The lower left corner is a mix of red, brown and purple colors. The red color is less distributed in the lower left corner. From the figure we can notice that the larger values of $\mu_{12}^M/\mu_B$ are concentrated in the upper right corner, which means that an increase in $g_{X}$ and $g_{YX}$ will promote its increase.

\section{Conclusion}

In this article, we first introduce $U(1)_X$SSM and then analyze the neutrino
transport magnetic moment on this basis. We study the transition magnetic moment of the Majorana neutrinos by applying the effective Lagrangian method and the on-shell scheme. We derive the Feynman diagrams and calculate the neutrino transport moment by combining the operators. We do a theoretical analysis of neutrino mixing.
Based on the five bounds of the neutrino experiment, we filter for the right effective light neutrino mass matrix element. Besides we perform a large number of numerical calculations and plot lines with different parameters versus $\mu_{ij}$ according to the experimental limits, followed by a large scan that yields rich numerical results. In the numerical calculation, at first we fit the experimental data on neutrino mass variance and mixing angle for the normal order condition. Then we select some sensitive parameters, including $g_X$,~$\lambda_H$,~$M_2$,~$\mu$ and $g_{YX}$. In the one dimensional plot, we analyze the parameters including $g_X$,~$\lambda_H$,~$M_2$ and~$\mu$ versus $\mu_{ij}^M/\mu_B$. In the scatter plot, we select three variants in TABLE \ref{c1} and study them. By analysing the numerical results, we understand the relationship between the selected parameters and $\mu_{ij}^M/\mu_B$, and they are indeed sensitive parameters.

Besides, we conclude that the order of magnitude of $\mu_{ij}^M/\mu_B$ is between $10^{-20}$ and $10^{-19}$.
From the diagrams, one can find that the numerical result of $\mu_{12}^M/\mu_B$ is at the order of $10^{-19}$.
The better limits on neutrino transition magnetic moment come from the recent XENONnT experiment\cite{XTnT}.
 We show the bounds at 90\% and 99\% C.L in the following table \ref{XT}.
 \begin{table}
\caption{The limits on neutrino transition magnetic moment at XENONnT experiment}
\begin{tabular}{|c|c|c|}
\hline
XENONnT  & 90\% C.L & 99\% C.L  \\
\hline
$|\mu_{12}/\mu_B| $ & $<6.77\times10^{-12}$ & $<9.63\times10^{-12}$   \\
\hline
$|\mu_{13}/\mu_B|$  & $<6.98\times10^{-12}$ & $<9.94\times10^{-12}$   \\
\hline
$|\mu_{23}/\mu_B| $ & $<9.04\times10^{-12}$ & $<12.9\times10^{-12}$   \\
\hline
\end{tabular}
\label{XT}
\end{table}
In the whole, the experimental sensitivity for $|\mu_{ij}/\mu_B|$ with $ i\neq j $ is a little smaller than $10^{-11}$.
In a Type-II radiative seesaw scenario\cite{TR}, the authors investigate neutrino magnetic moment,
and the obtained numerical results($|\mu_{ij}/\mu_B|$) are large and can reach $10^{-12}$.
Our corresponding results are at the order of $10^{-19}$, which are much smaller than their results\cite{TR}.

Compared with other conclusions\cite{13}, our results are two orders of magnitude larger than them.
The reason is that $U(1)_X$SSM has new gauge couplings $g_X$ and $g_{YX}$.
The vertices of $\nu_i$-$\chi_j^-$-$\tilde{e}_k^{*}$ are in Eq.(\ref{S1}).
On the face of it, Yukawa couplings($Y_{e,a}$ and $Y_{\nu,a}$) and gauge coupling $g_2$ are
obviously shown in the above equation. $Y_{\nu,a}$ are very tiny, and $Y_{e,a}$ are small.
Such as $Y_{e,2}$ for muon, with $\tan\beta=10$ the value of $Y_{e,2}$ is about 0.006, which is much smaller than $g_2$.
$g_X$ and $g_{YX}$ appear in the mass squared matrix of scalar lepton,
and their effects embody in the rotation matrix $Z^E$.
So the new gauge couplings $g_X$ and $g_{YX}$ can produce new effects.
Furthermore, the right-handed neutrinos and three Higgs singlets are added. They can produce new effects and improve the numerical results.

\begin{acknowledgments}

This work is supported by National Natural Science Foundation of China (NNSFC)(No.12075074),
Natural Science Foundation of Hebei Province(A2020201002, A2023201040, A2022201022, A2022201017, A2023201041),
Natural Science Foundation of Hebei Education Department (QN2022173),
Post-graduate's Innovation Fund Project of Hebei University (HBU2024SS042).
\end{acknowledgments}

\appendix
\section{}

The mass matrix for chargino reads:
\begin{eqnarray}
m_{\tilde{\chi}^-} = \left(
\begin{array}{cc}
M_2&\frac{1}{\sqrt{2}}g_2v_\mu\\
\frac{1}{\sqrt{2}}g_2v_d&\frac{1}{\sqrt{2}}\lambda_{H} v_S+\mu\end{array}
\right).
 \end{eqnarray}
This matrix is diagonalized by $U$ and $V$:\begin{eqnarray}
U^{*} m_{\chi^-} V^{\dagger}= m_{\chi^-}^{dia},
 \end{eqnarray}
with
\begin{eqnarray}
&&{\tilde{W}^-}=\sum_{t_2}U_{j1}^{*} \lambda_{j}^-,~~~~~~{\tilde{H}_{d}^-}=\sum_{t_2}U_{j2}^{*} \lambda_{j}^-,\nonumber\\
&&{\tilde{W}^+}=\sum_{t_2}V_{1j}^{*} \lambda_{j}^+,~~~~~~{\tilde{H}_{u}^-}=\sum_{t_2}V_{2j}^{*} \lambda_{j}^+.
 \end{eqnarray}

The mass matrix for slepton reads:
\begin{eqnarray}
m_{\tilde{e}}^2 = \left(
\begin{array}{cc}
m_{\tilde{e}_L\tilde{e}_L^{*}}&\frac{1}{2}({\sqrt{2}}{v_d}{T_e^\dagger}-{v_u}({\lambda_H}v_s+{\sqrt{2}\mu}){Y_e^\dagger})\\
\frac{1}{2}({\sqrt{2}}{v_d}{T_e}-{v_u}{Y_e}({\sqrt{2}\mu^{*}+v_s{\lambda_H}^{*}})&m_{\tilde{e}_R\tilde{e}_R^{*}}\end{array}
\right).
 \end{eqnarray}

\begin{eqnarray}
{m_{\tilde{e}_L\tilde{e}_L^{*}}}=&&M_{\tilde{L}}^2+ \frac{1}{8}\Big((g_{1}^{2} + g_{Y X}^{2})(-{v_u^2+v_d^2})+g_{YX}g_X(-2v_{\bar{\eta}}^2+2v_{\eta}^2-v_u^2+v_d^2)\nonumber\\&&+g_2^2(-v_d^2+v_u^2)\Big)
+\frac{1}{2}v_d^2{Y_e^\dagger}{Y_e},\\
{m_{\tilde{e}_R\tilde{e}_R^{*}}}=&&M_{\tilde{E}}^2-\frac{1}{8}\Big(2(g_{1}^{2} + g_{Y X}^{2})(-{v_u^2+v_d^2})+g_{YX}g_X(3v_d^2-3v_u^2-4v_{\bar{\eta}}^2+4v_{\eta}^2)\nonumber\\&&+g_X^2(-2v_{\bar{\eta}}^2
+2v_{\eta}^2-v_u^2+v_d^2)\Big)+\frac{1}{2}v_d^2{Y_e}{Y_e^\dagger}.
 \end{eqnarray}
This matrix is diagonalized by $Z^E$:
\begin{eqnarray}
Z^E{m_{\tilde{e}}}^2Z^{E,\dagger}=m_{2,\hat{e}}^{dia},
\end{eqnarray}
with
\begin{eqnarray}
&&{\tilde{e}_{L,i}}=\sum_{j}Z_{ji}^{E,*} \tilde{e}_j,~~~~~~{\tilde{e}_{R,i}}=\sum_{j}Z_{ji}^{E,*} \tilde{e}_j.\nonumber\\
 \end{eqnarray}

The mass matrix for neutrino reads:
\begin{eqnarray}
M_{\nu}=
\left({\begin{array}{*{20}{c}}
0 & \frac{\upsilon_u}{\sqrt{2}}Y_\nu^T  \\
\frac{\upsilon_u}{\sqrt{2}}Y_\nu & \sqrt{2}\upsilon_{\bar{\eta}}Y_X
\end{array}}
\right).
\end{eqnarray}

This matrix is diagonalized by $U_V$:
\begin{eqnarray}
U^{V,*}m_{\nu}U^{V,\dagger}=m_{\nu}^{dia},
\end{eqnarray}
with
\begin{eqnarray}
&&\nu_{L,i}=\sum_{j}U_{ji}^{V,*}\lambda_{\nu,j},~~~~~~\nu_{R,i}^*=\sum_{j}U_{ji}^{V}\lambda_{\nu,j}^*.\nonumber\\
 \end{eqnarray}

Here, we show some needed couplings in this model. We derive the vertices of
$\nu_i$-$\chi_j^-$-$\tilde{e}_k^{*}$
\begin{eqnarray}
&&\mathcal{L}_{\nu_i\chi_j^-\tilde{e}_k^{*}}=\bar{\nu}_i\Big\{\sum_{a=1}^{3}\Big(
U_{j2}^*U_{ia}^{V,*}Y_{e,a}Z_{k,3+a}^{E}-g_2U_{j1}^*U_{ia}^{V,*}Z_{ka}^{E}\Big)P_L,
\nonumber\\&&\hspace{1.8cm}+\sum_{a,b=1}^{3}Y_{\nu,ab}^{*}U_{i,3+a}^VZ_{kb}^{E}V_{j2}P_R\Big\}\chi_j^-\tilde{e}_k^{*}.
\label{S1}
\end{eqnarray}

\section{}
The expressions of $C_2^R, C_2^L, C_6^R, C_6^L$ are
\begin{eqnarray}
&&C_2^R=\sum\limits_{j=1}^{2}\sum\limits_{k=1}^{6}\vert{A_L^{ijk}}\vert^2F(X_{\tilde{L}_K},X_{X_j^\pm}),
\nonumber\\&&
C_2^L=\sum\limits_{j=1}^{2}\sum\limits_{k=1}^{6}\vert{A_R^{ijk}}\vert^2F(X_{\tilde{L}_K},X_{X_j^\pm}),
\nonumber\\
&&C_6^R=\sum\limits_{j=1}^{2}\sum\limits_{k=1}^{6}{A_R^{ijk}}{B_R^{ijk}} G(X_{\tilde{L}_K},X_{X_j^\pm}),
\nonumber\\&&
C_6^L=\sum\limits_{j=1}^{2}\sum\limits_{k=1}^{6}{A_L^{ijk}}{B_L^{ijk}} G(X_{\tilde{L}_K},X_{X_j^\pm}).
\end{eqnarray}
with the functions
\begin{eqnarray}
&&F(x,y)=\frac{1}{384 \pi ^2 \Lambda^2
   }\Big(\frac{ x^2-5 x y-2 y^2}{(x-y)^3}+\frac{6 x y^2
   (\log x-\log y)}{(x-y)^4}\Big),
\nonumber\\&&
G(x,y)=\frac{1}{64 \pi ^2 \Lambda^2}\frac{x^2-y^2+2 x y (\log
   y- \log x)}{(x-y)^3}.
\end{eqnarray}
The used couplings are
\begin{eqnarray}
&&A_L=\sum\limits_{a=1}^{3}(U_{j2}^*
U_{ia}^{V,*}Y_{e,a}Z_{k3+a}^E-g_2U_{j1}^*
U_{ia}^{V,*}Z_{ka}^E),
\nonumber\\&&
A_R=\sum\limits_{a,b=1}^{3}Y_{\nu,ab}^*U_{i3+a}^VZ_{kb}^EV_{j2},
\nonumber\\&&
B_R=A_L^*,~~~~~~ B_L=A_R^*.
\end{eqnarray}

\section{}
The effective light neutrino mass matrix can be written as
\begin{eqnarray}
\mathcal{M}_\nu^{eff}\approx -\Big(\frac{v_uY_{\nu}}{\sqrt{2}}\Big)
\Big( \sqrt{2}v_{\bar{\eta}}Y_{X}\Big)^{-1}\Big(\frac{v_uY_{\nu}}{\sqrt{2}}\Big)^T.
\end{eqnarray}
Using the " top-down " method\cite{TOP},  we get the Hermitian matrix
\begin{eqnarray}
{\cal H}=(\mathcal{M}_\nu^{eff})^{\dagger}\mathcal{M}_\nu^{eff}.
\end{eqnarray}
Besides we can diagonalize the $3\times3$ matrix ${\cal H}$ to gain three eigenvalues
\begin{eqnarray}
&&m_1^2={a\over3}-{1\over3}p(\cos\phi+\sqrt{3}\sin\phi),
\nonumber\\
&&m_2^2={a\over3}-{1\over3}p(\cos\phi-\sqrt{3}\sin\phi),
\nonumber\\
&&m_3^2={a\over3}+{2\over3}p\cos\phi.\label{massQ}
\end{eqnarray}
These parameters can be given by
\begin{eqnarray}
&&p=\sqrt{a^2-3b}, ~~~~~\phi={1\over3}\arccos({1\over p^3}(a^3-{9\over2}ab+{27\over2}c)),
~~~~a={\rm Tr}({\cal H}),\nonumber\\
&&b={\cal H}_{11}{\cal H}_{22}+{\cal H}_{11}{\cal H}_{33}+{\cal H}_{22}{\cal H}_{33}
-{\cal H}_{12}^2 -{\cal H}_{13}^2-{\cal H}_{23}^2,~~~~c={\rm Det}({\cal H}).
\end{eqnarray}
We take the normal ordering (NO), so:
\begin{eqnarray}
&&m_{\nu_1}<m_{\nu_2}<m_{\nu_3}, ~~~m_{\nu_1}^2=m_1^2,\quad m_{\nu_2}^2=m_2^2,\quad m_{\nu_3}^2=m_3^2, \nonumber\\
&&\Delta m_{\odot}^2 = m_{\nu_2}^2-m_{\nu_1}^2 ={2\over \sqrt{3}}p\sin\phi>0,\nonumber\\
&&\Delta m_{A}^2 =m_{\nu_3}^2-m_{\nu_1}^2 =p(\cos\phi+{1\over\sqrt{3}}\sin\phi)>0.
\end{eqnarray}

From the mass squared matrix ${\cal H}$, one gets
the normalized eigenvectors
\begin{eqnarray}
&&\left(\begin{array}{c}\Big(U_\nu\Big)_{11}\\
\Big(U_\nu\Big)_{21}\\\Big(U_\nu\Big)_{31}
\end{array}\right)={1\over\sqrt{|X_1|^2+|Y_1|^2+|Z_1|^2}}\left(\begin{array}{c}
X_1\\Y_1\\Z_1\end{array}\right), \nonumber\\
&&\left(\begin{array}{c}\Big(U_\nu\Big)_{12}\\
\Big(U_\nu\Big)_{22}\\\Big(U_\nu\Big)_{32}
\end{array}\right)={1\over\sqrt{|X_2|^2+|Y_2|^2+|Z_2|^2}}\left(\begin{array}{c}
X_2\\Y_2\\Z_2\end{array}\right),
\nonumber\\
&&\left(\begin{array}{c}\Big(U_\nu\Big)_{13}\\
\Big(U_\nu\Big)_{23}\\\Big(U_\nu\Big)_{33}
\end{array}\right)={1\over\sqrt{|X_3|^2+|Y_3|^2+|Z_3|^2}} \left(\begin{array}{c}
X_3\\Y_3\\Z_3\end{array}\right).
\end{eqnarray}

The concrete forms of $X_I,Y_I,Z_I$ for $I=1,2,3$ are shown here
\begin{eqnarray}
&&X_1=({\cal H}_{22}-m_{{\nu_1}}^2)({\cal H}_{33}-m_{{\nu_1}}^2)-{\cal H}_{23}^2, ~~~~Y_1={\cal H}_{13}{\cal H}_{23}-{\cal H}_{12}({\cal H}_{33}-m_{{\nu_1}}^2), \nonumber\\
&&Z_1={\cal H}_{12}{\cal H}_{23}-{\cal H}_{13}({\cal H}_{22}-m_{{\nu_1}}^2),
~~~~~~~~~~X_2={\cal H}_{13}{\cal H}_{23}-{\cal H}_{12}\Big({\cal H}_{33}-m_{{\nu_2}}^2\Big),
\nonumber\\
&&Y_2=({\cal H}_{11}-m_{{\nu_2}}^2)({\cal H}_{33}-m_{{\nu_2}}^2)-{\cal H}_{13}^2,
~~~~Z_2={\cal H}_{12}{\cal H}_{13}-{\cal H}_{23}\Big({\cal H}_{11}-m_{{\nu_2}}^2\Big),
\nonumber\\
&&X_3={\cal H}_{12}{\cal H}_{23}-{\cal H}_{13}\Big({\cal H}_{22}-m_{{\nu_3}}^2\Big),
~~~~~~~~~Y_3={\cal H}_{12}{\cal H}_{13}-{\cal H}_{23}\Big({\cal H}_{11}-m_{{\nu_3}}^2\Big),
\nonumber\\
&&Z_3=({\cal H}_{11}-m_{{\nu_3}}^2)({\cal H}_{22}-m_{{\nu_3}}^2)-{\cal H}_{12}^2.
\end{eqnarray}
The mixing angles among three tiny neutrinos can be defined as follows
\begin{eqnarray}
&&\sin\theta_{13}=\Big|\Big(U_\nu\Big)_{13}\Big|,~~~~~~~~~~~~~~~\cos\theta_{13}=\sqrt{1-\Big|\Big(U_\nu\Big)_{13}\Big|^2},\nonumber\\
&&
\sin\theta_{23}={\Big|\Big(U_\nu\Big)_{23}\Big|\over\sqrt{1-\Big|\Big(U_\nu\Big)_{13}\Big|^2}},~~~~~~
\cos\theta_{23}={\Big|\Big(U_\nu\Big)_{33}\Big|\over\sqrt{1-\Big|\Big(U_\nu\Big)_{13}\Big|^2}},\nonumber\\
&& \sin\theta_{12}={\Big|\Big(U_\nu\Big)_{12}\Big|\over\sqrt{1-\Big|\Big(U_\nu\Big)_{13}\Big|^2}},~~~~~~
\cos\theta_{12}={\Big|\Big(U_\nu\Big)_{11}\Big|\over\sqrt{1-\Big|\Big(U_\nu\Big)_{13}\Big|^2}}.
\end{eqnarray}


\begin{thebibliography}{0}

\bibitem{b0}S.L. Glashow, H. Georgi, Phys. Rev. Lett. {\bf 32} (1974) 438-441.
\bibitem{b1}S. Weinberg, Phys. Rev. Lett. {\bf 19} (1967) 1264-1266.
\bibitem{b2}S. Weinberg, Phys. Rev. D. {\bf 19} (1979) 1277-1280.
\bibitem{b3}A. Salam, J. C. Ward, Phys. Rev. Lett. {\bf 30} (1973) 1268-1271.


\bibitem{n0}H.P. Nilles, Phys. Rept. {\bf 110} (1984) 1-162.
\bibitem{n1}H.E. Haber, G.L. Kane, Phys. Rept. {\bf 117} (1985) 75-263.
\bibitem{n2} Rosiek, Phys. Rev. D {\bf 41} (1990) 3464.


\bibitem{31}U. Ellwanger, C. Hugonie, A.M. Teixeira, Phys. Rep. {\bf 496} (2010) 1-77.
\bibitem{32}B. Yan, S.M. Zhao, T.F. Feng, Nucl. Phys. B {\bf 975} (2022) 115671.





\bibitem{13}H.B. Zhang, T.F. Feng, Zhao Feng Ge, et al., JHEP {\bf 02}(2014) 012.

\bibitem{111}A.D. Gouvea, S. Shalgar, JCAP {\bf 04} (2013) 018.

\bibitem{112}V. Brdar, A. Greljo, J. Kopp, et al., JCAP {\bf 01} (2021) 039.


\bibitem{7}S.M. Zhao, T.F. Feng, M.J. Zhang, et al., JHEP {\bf 02} (2020) 130.


\bibitem{UMSSM5}G. Belanger, J.D. Silva, H. M. Tran, Phys. Rev. D {\bf 95} (2017) 115017.
\bibitem{B-L1}V. Barger, P.~F.~Perez, S.~Spinner, Phys. Rev. Lett. {\bf 102} (2009) 181802.
\bibitem{B-L2}P. H. Chankowski, S.~Pokorski, J.~Wagner, Eur. Phys. J. C {\bf 47} (2006) 187.
\bibitem{gaugemass}J. L. Yang, T.~F.~Feng, S.~M.~Zhao, et al., Eur. Phys. J. C {\bf 78} (2018) 714.


\bibitem{Feng}T.F. Feng, L.~Sun, X.Y.~Yang, Nucl. Phys. B {\bf 800} (2008) 221.



\bibitem{18}R.L. Workman, et al. (Particle Data Group), Prog. Theor. Exp. Phys. {\bf 2022} (2022) 083C01.


\bibitem{su1}CMS Collaboration, Phys. Lett. B {\bf 716} (2012) 30.
\bibitem{su2}ATLAS Collaboration, Phys. Lett. B {\bf 716} (2012) 1.

\bibitem{1251} M. Carena, J.R. Espinosa, M. Quir\'{o}s, et al., Phys. Lett. B {\bf 355} (1995) 209.
 \bibitem{1252}  M. Carena, S. Gori, N.R. Shah, et al., JHEP {\bf 03} (2012) 014.

\bibitem{1}F.P. An et al. Phys. Rev. Lett. {\bf 130} (2023) 161802.
\bibitem{2}R. Abbasi et al. Phys. Rev. D  {\bf 108} (2023) 012014.
\bibitem{3}K. Abe et al. Phys. Rev. D   {\bf 108} (2023) 072011.

\bibitem{21}Marcela Carena1, Alejandro Daleo, et al. Phys. Rev. D {bf 70} (2004) 093009.

\bibitem{XTnT} E. Aprile et al., (XENON), Phys. Rev. Lett. \textbf{129} (2022) 161805.

\bibitem{TR}S. Singirala, D.K. Singha, R. Mohanta, Phys. Rev. D \textbf{109} (2024) 075031.

\bibitem{TOP} B.Dziewit, S.Zajac, M. Zralek, Acta Phys. Pol. B \textbf{42} (2011) 2509.

\end{thebibliography}
\end{document}